\documentclass{nature}

\usepackage{epsfig,graphicx}
\usepackage{amssymb,amsmath}
\usepackage{latexsym}
\usepackage{enumitem}
\usepackage{afterpage}  

\usepackage{multibib}

\usepackage{bm}
\newcommand*{\B}[1]{\ifmmode\bm{#1}\else\textbf{#1}\fi}

\makeatletter
\let\saved@includegraphics\includegraphics
\AtBeginDocument{\let\includegraphics\saved@includegraphics}
\renewenvironment*{figure}{\@float{figure}}{\end@float}
\makeatother

\newcites{main,methods}{{},{}}

\usepackage{geometry}
\title{Strong disc winds traced throughout outbursts in black-hole X-ray binaries}

\author{B.E. Tetarenko*$^{1}$, J.-P. Lasota$^{2,3}$, C.O. Heinke$^{1}$, G. Dubus$^{4}$, and G.R. Sivakoff$^{1}$}

\begin{document}

\maketitle

\begin{affiliations}
 \item Department of Physics, University of Alberta, CCIS 4-181, Edmonton, AB T6G 2E1, Canada
 \item Institut d'Astrophysique de Paris, CNRS et Sorbonne Universit\'es, UPMC Paris 06, UMR 7095, 98bis Bd Arago, 75014 Paris, France
 \item Nicolaus Copernicus Astronomical Centre, Polish Academy of Sciences, ul. Bartycka 18, 00-716 Warsaw, Poland
 \item Univ. Grenoble Alpes, CNRS, Institut de Plan\'etologie et d'Astrophysique de Grenoble (IPAG),F-38000, Grenoble, France
\end{affiliations}

\begin{abstract}
Recurring outbursts associated with matter flowing onto compact stellar remnants (black-holes, neutron stars, white dwarfs) in close binary systems,
provide strong test beds for constraining the poorly understood accretion process. The efficiency of angular momentum (and thus mass) transport in accretion discs, which has traditionally been encoded in the $\B{\alpha}$-viscosity parameter, shapes the light-curves of these outbursts. 
Numerical simulations of the magneto-rotational instability that is believed to be the physical mechanism behind this transport
find values of $\B{\alpha \sim0.1-0.2$}\cite{hirose2014,coleman2016,scepi2017}, as required from observations of accreting white dwarfs\cite{kotko2012}.
Equivalent $\B{\alpha}$-viscosity parameters have never been estimated in discs around neutron stars or black holes.
Here we report the results of an analysis of archival X-ray light-curves of twenty-one black hole X-ray binary outbursts. Applying a Bayesian approach for a model of accretion allows us to determine corresponding $\B{\alpha}$-viscosity parameters, directly from the light curves, to be $\B{\alpha}\sim$0.2--1. 
This result may be interpreted either as a strong intrinsic rate of angular momentum transport in the disc, which can only be sustained by the magneto-rotational instability if a large-scale magnetic field threads the disc \cite{lesur2013,bai2013,salvesen2016}, or as a direct indication that mass is being lost from the disc through substantial mass outflows strongly shaping the X-ray binary outburst. 
Furthermore, the lack of correlation between our estimates of $\B{\alpha}$-viscosity and accretion state implies that such outflows can remove a significant fraction of disc mass in all black hole X-ray binary accretion states, favouring magnetically-driven winds over thermally-driven winds that require specific radiative conditions\cite{higginbottom2015}. 
\end{abstract}

The disc-instability model\cite{osaki1974,meyer81,Smak84,FLP83} was developed to explain outbursts in compact binaries where a white dwarf accretes from a low-mass companion.
A cool (neutral) quiescent disc is built up through steady mass transfer from the companion star, causing the disc temperature to rise. Eventually at some radius (called the ignition radius) the disc temperature will reach the temperature where hydrogen ionizes. This triggers a thermal-viscous instability within the disc due to the steep temperature dependence of opacity in this temperature range. As a result, the disc will cycle between a hot, ionized, outburst state and a cold, neutral, quiescent state. 
The growth of the thermal-viscous instability at the ignition radius results in two heating fronts propagating inwards and outwards through the disc. This brings the disc into a hot state causing rapid in-fall of matter onto the compact object, and a bright optical and UV outburst. 

As the disc is depleted over time (because mass falls onto the compact stellar remnant at a higher rate than it is being transferred from the companion star), the temperature and mass accretion rate in the outer radii will eventually be reduced to the point where hydrogen can recombine. This triggers the formation and propagation of a cooling front that returns the disc to its quiescent (neutral) state. While this predicted behaviour, characterized by alternating periods of disc-outbursts and quiescence, matches observations of accreting white dwarfs well, changes are needed for close binaries containing more compact stellar remnants (neutron stars and stellar-mass black holes) called low-mass X-ray binaries. 

There are 18 confirmed black-hole low-mass X-ray binaries in our Galaxy, identified through bright X-ray outbursts indicating rapid accretion episodes \cite{tetarenkob2015}. 
These outbursts\cite{tetarenkob2015} last considerably longer, and recur significantly less frequently, than in many types of accreting white dwarfs, due to heating of the outer disc by X-rays emitted in the inner regions of the accretion flow. \cite{vanp96}. 

X-ray irradiation will keep the accretion disc in its hot (ionized) state over the viscous timescale. This timescale, which is encoded in observed outburst light-curves,
is directly related to the efficiency of angular momentum transport, and thus, provides a means to measure this efficiency. See Methods and Extended Data Figure 1 for a detailed discussion of the characteristic three-stage outburst decay profile present in a black-hole low-mass X-ray binary light-curve.

The magneto-rotational instability is thought to provide the physical mechanism behind angular momentum (and mass) transport in accretion discs\cite{balhaw98}. The effective viscosity in these discs, commonly parametrized using the $\alpha$-viscosity prescription\cite{shak73}, encapsulates the efficiency of this transport process.
Physically, the $\alpha$-viscosity parameter sets the viscous time of the accretion flow through the disc, and thus, according to the disc-instability picture, is encoded within the decay profile of an outburst light-curve. A disc with higher viscosity (higher $\alpha$) in outburst will accrete mass more quickly, resulting in shorter decay times and shorter outburst durations\cite{dubus2001}.

The $\alpha$-viscosity has only ever been inferred in (non-irradiated) discs around accreting white dwarfs by comparing the outburst timescales from observed light-curves to synthetic model light-curves created by numerical disc codes for different $\alpha$-viscosity inputs\cite{kotko2012}. $\alpha$-viscosity parameters have never before been measured in irradiated accretion discs, such as those around stellar-mass black holes in low-mass X-ray binaries. The assertion\cite{KPL07} that $\alpha \simeq 0.2$--0.4 in such systems, was deduced from the calculations\cite{dubus2001} of ``detailed models of complete light curves'', but not from detailed comparison of models with observations. Note that we learned of the recent study of black hole low-mass X-ray binary 4U1543-475\cite{lipunova2017} after acceptance of this manuscript.

Accordingly, we have built a novel Bayesian approach that characterizes the angular-momentum (and mass) transport occurring in the discs in low-mass X-ray binary systems.
The $\alpha$-viscosity parameter in a hot, outbursting disc ($\alpha_h$) sets the timescale on which matter moves through the hot (ionized) portion of the disc, and thus, controls the duration of the first stage of the decay profile observed in an X-ray light-curve (see Methods for details). This (viscous) timescale will vary depending upon the mass of the compact object and the size of the accretion disc, where the size of the disc itself is governed by the ratio of component masses in the system and the binary orbital period. 
To reconcile the multi-level, interconnected relationships existing between these parameters defining the properties of the accretion flow, we use a powerful statistical data analysis technique referred to as Bayesian hierarchical modelling.
This allows us to derive: (i) timescales associated with individual stages of the outburst decay, and (ii) the rate of mass accretion through the disc during, and the time of occurrence of, the transitions between the individual decay stages.
Ultimately, this Bayesian technique allows us to effectively take into account our prior knowledge of the orbital parameters that define a low-mass X-ray binary system (black-hole mass, companion mass, and orbital period), to sample the $\alpha$-viscosity parameter in a hot, outbursting disc ($\alpha_h$) directly from its observed X-ray light-curve. For details of this Bayesian methodology, see the Methods section.

We analyzed a representative sample of X-ray light-curves of 21 individual outbursts of 12 black-hole low-mass X-ray binary systems, from the WATCHDOG project\cite{tetarenkob2015} (Extended Data Table 1).
Figure 1 shows examples of the analytical irradiated disc-instability model fit to observed data.  In this figure, we overlay predicted decay profiles that illustrate how varying the $\alpha$-viscosity parameter in a hot, outbursting disc changes the predicted light-curve decay profile.   
For these 21 outbursts, we derive $0.19<\alpha<0.99$ (see Figure 2 and Extended Data Table 2). These results represent the first-ever derivation of $\alpha$-viscosity parameters in low-mass X-ray binary accretion discs from a fit to the observed outburst light-curves of such systems.

There are two probable explanations for the 
high values of $\alpha$-viscosity we measure in black-hole low-mass X-ray binary discs.
The first is that we are actually measuring the intrinsic $\alpha$-viscosity parameters of these discs.
The only way to reproduce such high intrinsic $\alpha$-viscosity parameters in accretion disc simulations is for a net magnetic field to thread the disc, with concurrent mass outflows strongly shaping the outburst as a whole.
Simulations of angular momentum transport driven by the magneto-rotational instability, carried out in vertically-stratified boxes representing a local patch of the disc (shearing box), typically yield $\alpha\approx 0.02$ without a net magnetic flux \cite{davis2010,simon2012}. 
Convection enhances transport to $\alpha\approx 0.2$ in the conditions appropriate to accreting white dwarfs  \cite{hirose2014,coleman2016,scepi2017}. This is consistent with the values deduced from observations of outbursts of the non-irradiated discs around these objects \cite{kotko2012}, but is insufficient to explain the higher values of $\alpha\gtrsim 0.2$ that we measure from black hole outbursts. 
However, when the shearing box is threaded by a net magnetic flux, simulations show that $\alpha$ scales as $\beta^{-1/2}$, where $\beta$ is the ratio of the thermal pressure to the imposed magnetic pressure, reaching values as high as $\alpha \approx 1$ when $1\lesssim \beta\lesssim 10^3$, with $\beta=1$ the lower limit for the magneto-rotational instability to operate in a thin disc  \cite{lesur2013,bai2013,salvesen2016}. Hence, strong intrinsic angular momentum transport indicates the presence of a large-scale field in the accretion disc, whose origin and evolution have yet to be determined in black hole transients.
Moreover, simulations reproducing high intrinsic $\alpha$-viscosity also display strong outflows, which actually do not remove much angular momentum, thus angular momentum transport is still primarily driven by the magneto-rotational instability.

The second possibility is that the intrinsic $\alpha$-viscosity parameters in black-hole low-mass X-ray binary accretion discs are smaller than we measure (e.g., comparable to $\alpha\sim0.2$), and unspecified strong mass outflows are significantly shaping the overall observed light-curve profiles.
Figure 3 illustrates how including a mass (and angular momentum) loss term within the irradiated disc instability model mimics the effect that a high $\alpha$-viscosity has on the light-curve decay profile.

In both cases, significant outflows appear to 
play a key role in regulating the disc-accretion process. 
Strong mass outflows have been observed in outbursting low-mass X-ray binaries in the soft/intermediate states or at high flux levels ($ > 10$\% Eddington) in the hard state  \cite{miller06,ponti2012,neilsen2013} in the form of accretion disc winds.
These outflows have been attributed to thermal winds driven by X-ray irradiation or to magnetic winds driven by centrifugal acceleration along magnetic field lines anchored in the disc \cite{ohsuga2011,higginbottom2015}.
It has recently been shown that thermally-driven winds (e.g., Compton-heated winds\cite{ponti2012}), can only be produced in the soft accretion state, as the ionization state of the wind becomes unstable in the hard state (e.g. \cite{chakravorty2013,bianchi2017}).
The absence of correlation between the values of $\alpha$ and the X-ray flux or accretion state in our outburst sample, suggests that the outflow mechanism is generic, and favours magnetically-driven over thermally-driven outflows.

\nocitemain{hirose2014,coleman2016,scepi2017,kotko2012,lesur2013,bai2013,salvesen2016,higginbottom2015,warner1995,tetarenkob2015,osaki1974,meyer81,Smak84,FLP83,vanp96,kingrit8,dubus1999,dubus2001,balhaw98,shak73,KPL07,davis2010,simon2012,miller06,ponti2012,neilsen2013,ohsuga2011,king98,chakravorty2013,bianchi2017,lipunova2017}
\bibliographystylemain{naturemag}
\bibliographymain{nature_refs.bib}


\begin{addendum}
 \item B.E.T. would like to thank participants of the ``disks17: Confronting MHD Theories of Accretion Disks with Observations'' program, held at the Kavli Institute for Theoretical Physics (KITP), for their feedback and useful comments on this project. Especially, Alexandra Veledina for advice given regarding the X-ray light curve analysis, and Phil Charles for his comments on the manuscript. B.E.T., G.R.S., and C.O.H. acknowledge support by NSERC Discovery Grants, and C.O.H. by a Discovery Accelerator Supplement. This research was supported in part by the National Science Foundation under Grant No. NSF PHY-1125915, via support for KITP. J.-P.L acknowledges support by the Polish National Science Centre OPUS grant 2015/19/B/ST9/01099. J.-P.L and G.D. also acknowledge support from the French Space Agency CNES. This research has made use of data, software, and/or web tools obtained from the High Energy Astrophysics Science Archive Research Center (HEASARC), a service of the Astrophysics Science Division at NASA/GSFC and of the Smithsonian Astrophysical Observatory's High Energy Astrophysics Division, data supplied by the UK Swift Science Data Centre at the University of Leicester, and data provided by RIKEN, JAXA, and the MAXI team. This work has also made extensive use of NASA's Astrophysics Data System (ADS).
 \item[Author Contribution] B.E.T performed the analysis of the X-ray data, wrote the Markov-Chain Monte-Carlo light-curve fitting algorithm and performed the light-curve fitting, built the Bayesian hierarchical methodology, and wrote the paper. J.-P.L. helped formulate the analytical version of the irradiated-disc instability model that was fit to the X-ray light-curves, contributed to interpretation of the data, and assisted in writing the discussion in the paper. C.O.H. assisted in the analysis of the X-ray data and the light-curve fitting process, and contributed to the interpretation of the data. G.D. contributed to the interpretation of the data, and assisted in writing the discussion in the paper. G.R.S. assisted in writing the paper and contributed to the interpretation of the data.
 \item[Author Information] Reprints and permissions information is available at www.nature.com/reprints. The authors declare no competing financial interests. Correspondence and requests for materials should be addressed to BET (btetaren@ualberta.ca).

\end{addendum}



\begin{figure}%
\resizebox{\textwidth}{!}{%
\includegraphics[width=\columnwidth]{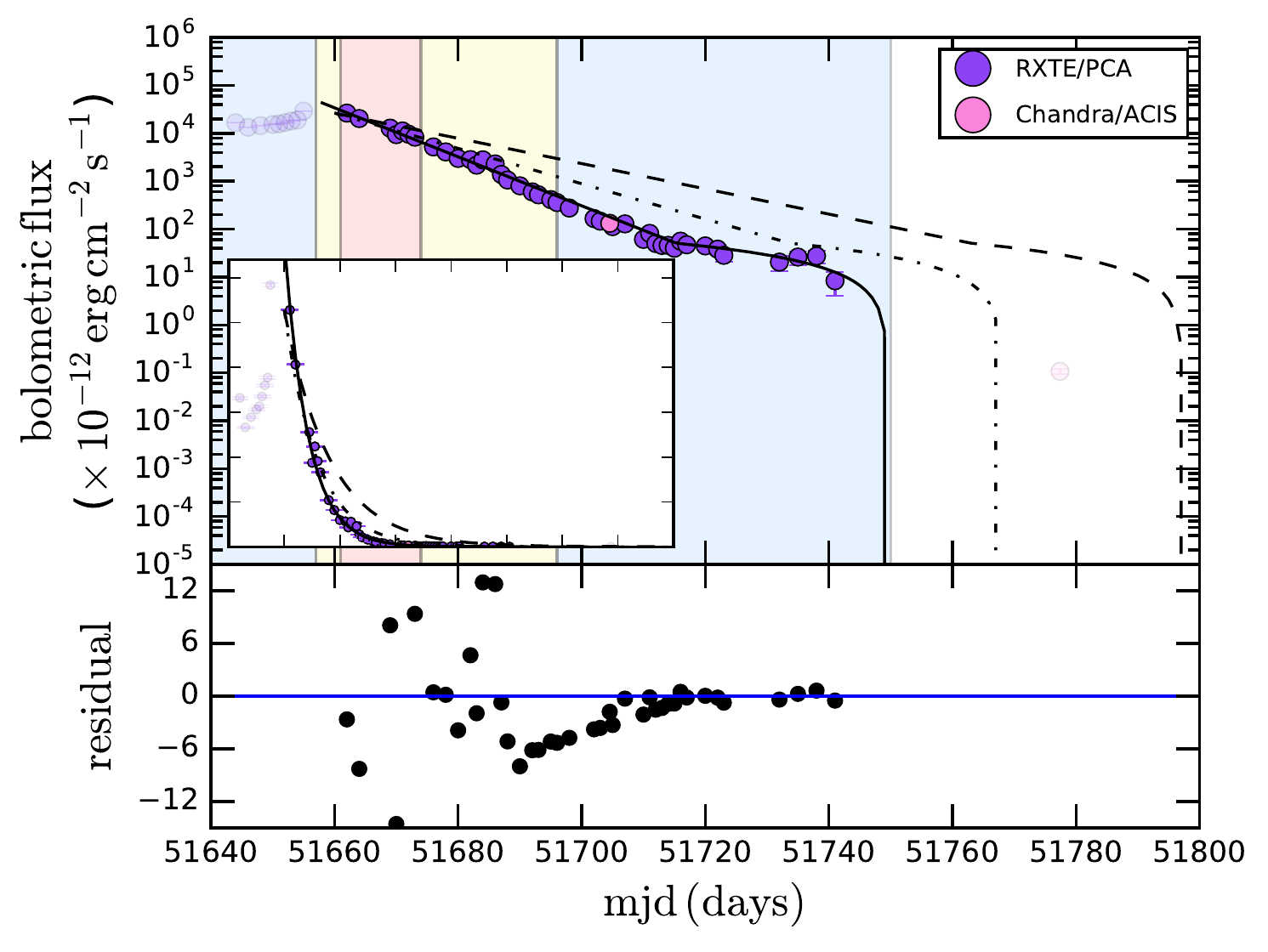}
\includegraphics[width=\columnwidth]{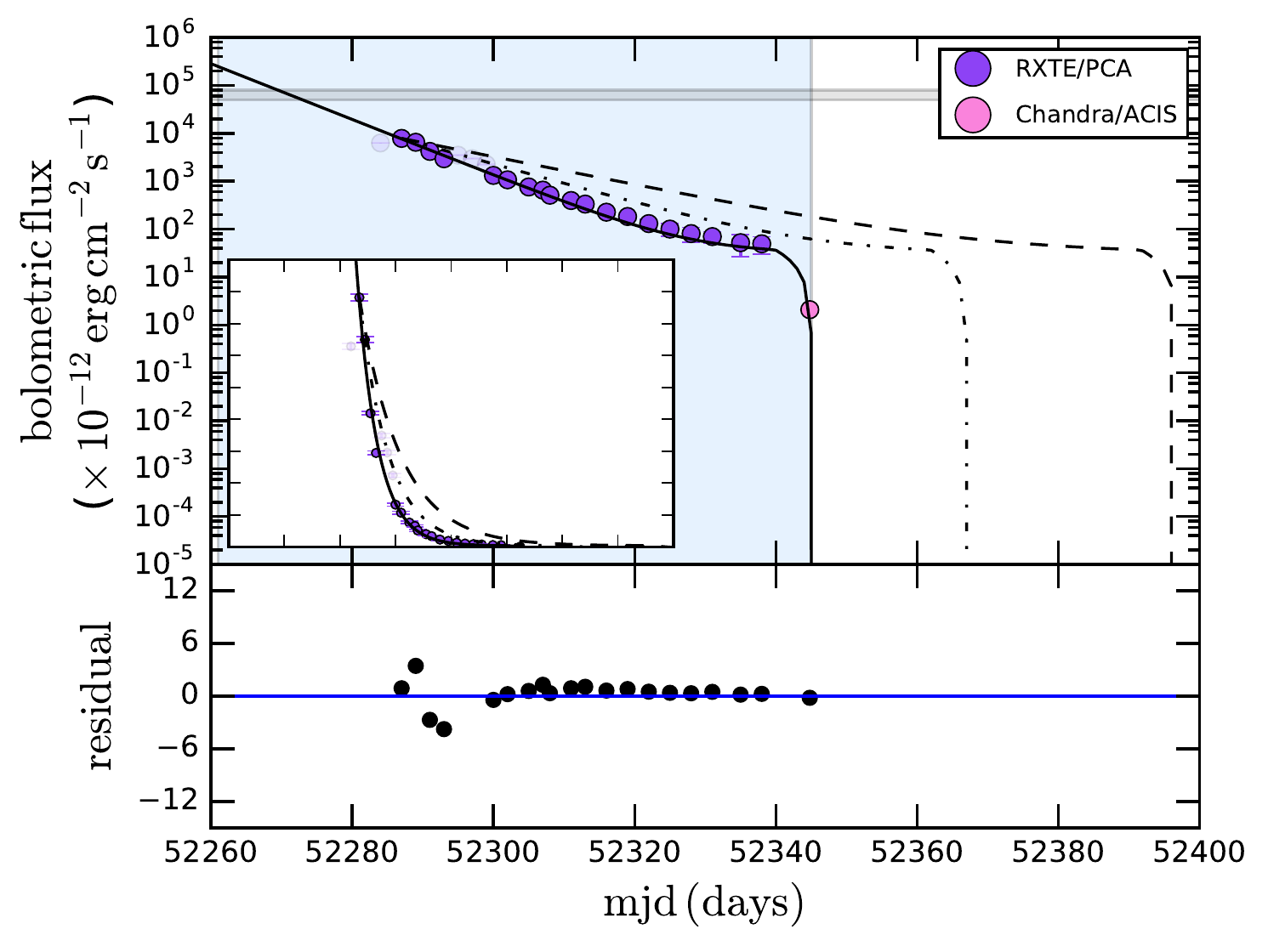}
}
\caption{Example low-mass X-ray binary outburst light-curves. The figure displays the observed X-ray light-curves for the (left panel) 2000 and (right panel) 2001/2002 outbursts of low-mass X-ray binary XTE J1550$-$564, which harbours a $10.4\pm2.3M_{\odot}$ black-hole\cite{tetarenkob2015}. In this source, which has undergone multiple outbursts in the past two decades, we measure an extremely high $\alpha$-viscosity.  
Error bars show the instruments' statistical uncertainties.
Shaded colours show the accretion state(s) of the source during the outbursts: blue = hard, yellow = intermediate, red = soft. Although XTE J1550$-$564 transitions from the soft to hard accretion states during the decay of the 2000 outburst, the light-curve shows  no signature of that transition. Disc outflows have only been observed in the soft/intermediate states or at high flux levels ($>10$\% Eddington)\cite{neilsen2013}; above the grey region (left panel) in the hard state. Coloured circles represent data from different X-ray instruments;  translucent data indicate the rise of the outburst, which was not included in the fits. The inset axes shows the outburst on a linear scale. The best fit analytical model (solid black line) and residuals (lower panel) are displayed in both figures. We measure $\alpha=0.96\pm0.15$ and $\alpha=0.99^{+0.15}_{-0.14}$ from the 2000 and 2001/2002 outburst light-curves, respectively. We over-plot the resulting decay profiles corresponding to $\alpha=0.7$ (dot-dashed line) and $\alpha=0.5$ (dashed line), demonstrating how the light-curve shape changes with different values of the $\alpha$-viscosity parameter.}
\label{fig:example_lcs}
\end{figure}

\clearpage

\begin{figure}
  \center
\resizebox{89mm}{!}{
\includegraphics[width=\columnwidth]{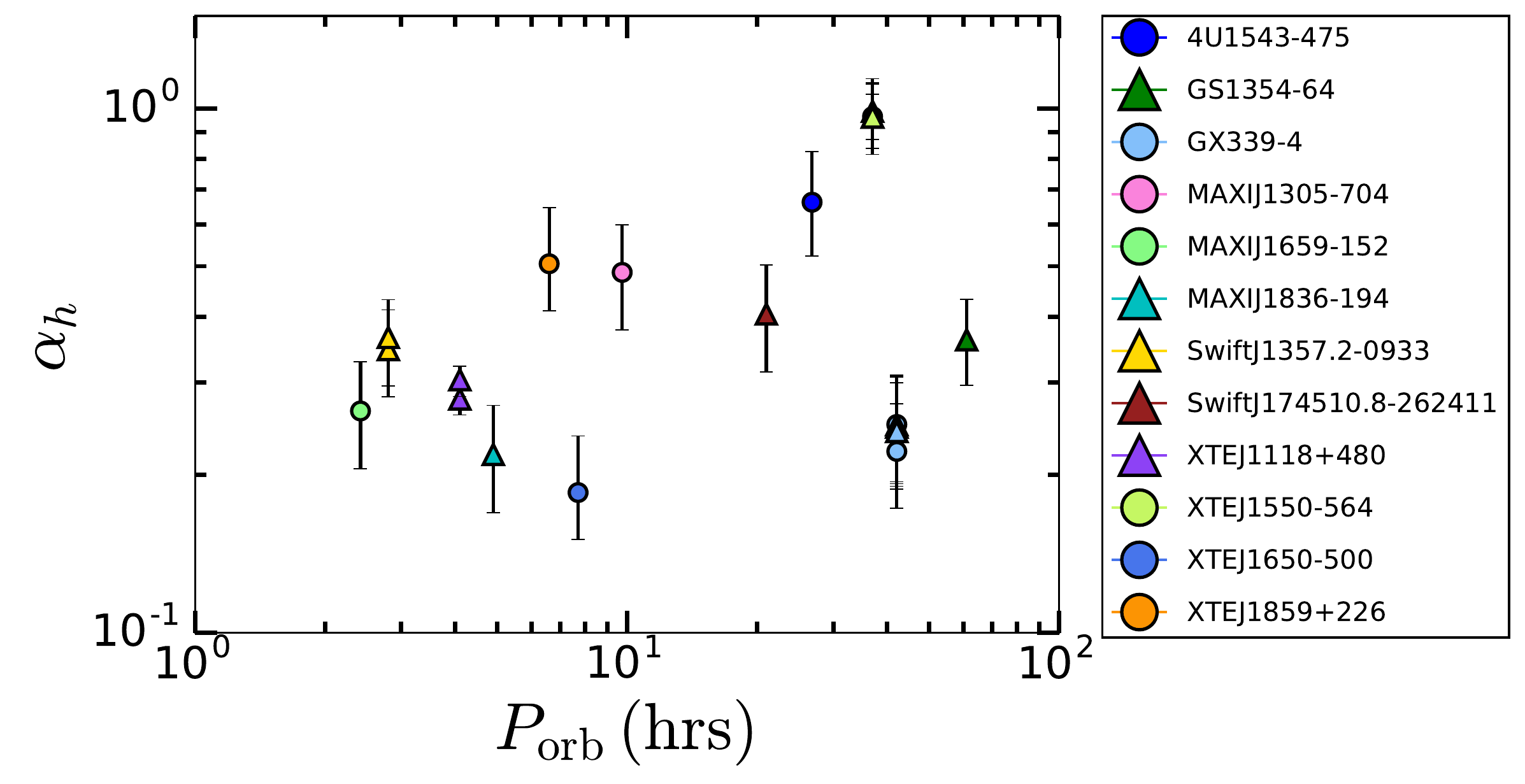}}
\caption{Characterization of the mass transport process at work in accretion discs.
$\alpha$-viscosity, the parameter which encompassed the efficiency of angular-momentum (and mass) transport in accretion discs, as derived by our Bayesian methodology, is plotted vs.\ binary system orbital period ($P_{\rm orb}$), for the 21 individual outbursts occurring in our sample of 12 Galactic black hole low-mass X-ray binaries with measured orbital periods.  The different colours represent individual sources. The shapes indicate accretion state(s) reached by the source during outburst: (circles) hard/intermediate/soft states reached and (triangles) only hard state reached.Error bars represent the 68\% confidence interval. $\alpha$-viscosity parameters are derived in both outbursts where the source cycles through all the accretion states and those where the source remains only in the hard state.
}
\label{fig:alpha_porb}
\end{figure}

\clearpage

\begin{figure}
  \center
\resizebox{89mm}{!}{
\includegraphics[width=\columnwidth]{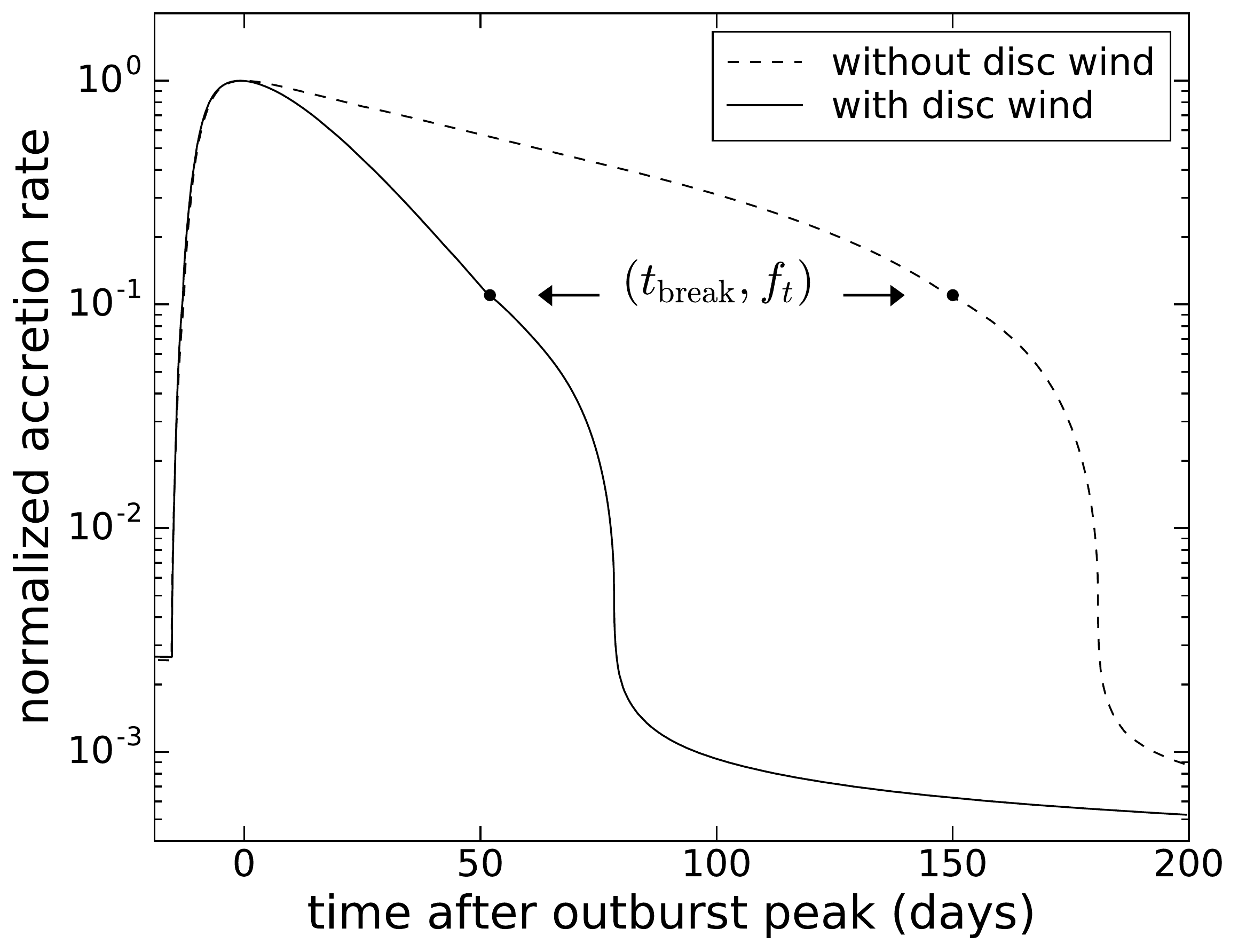}}
\caption{The ``disc wind'' toy model. Two model light-curves for an irradiated disc around a $6M_{\odot}$ black-hole and an $\alpha=0.2$ accretion disc are shown: (dashed line) no mass loss present, and (solid line) including a mass loss term during outburst. The latter is computed assuming mass loss is proportional to central mass-accretion rate onto the black-hole ($\dot{M_w}=\epsilon_w \dot{M_c}$) during the decay (meant to be representative of a disc ``wind'' type outflow). While the profile shape remains the same, the effective timescale ($\tau_e$) is reduced to $(1-\epsilon_w)\tau_e$. Thus, as fraction of mass lost increases, $\tau_e$ decreases, mimicking the effect an arbitrary large $\alpha$-viscosity parameter has on the light-curve profile (i.e., a high $\alpha$-viscosity parameter corresponds to a fast decay).
A measurement of $\alpha=1$ would correspond to a disc with $\alpha=0.2$, $\epsilon_w=0.8$ in the toy model, indicative of a significant outflow. Note that, while this model assumes the local outflow rate is related to the local accretion rate in the disc, this need not be the case. Further, this simplifying assumption, used purely to numerically solve the light-curve, will limit what we can say on how much mass is lost in the outflow. While this model requires $\dot{M_w}/\dot{M_c}<1$, it is certainly possible that the outflow rate is larger than the central mass accretion rate onto the black-hole.  }
\label{fig:toy_model}
\end{figure}

\clearpage

\setcounter{figure}{0}
\renewcommand{\figurename}{Extended Data Figure}
\begin{figure}
  \center
\resizebox{89mm}{!}{
\includegraphics[width=\columnwidth]{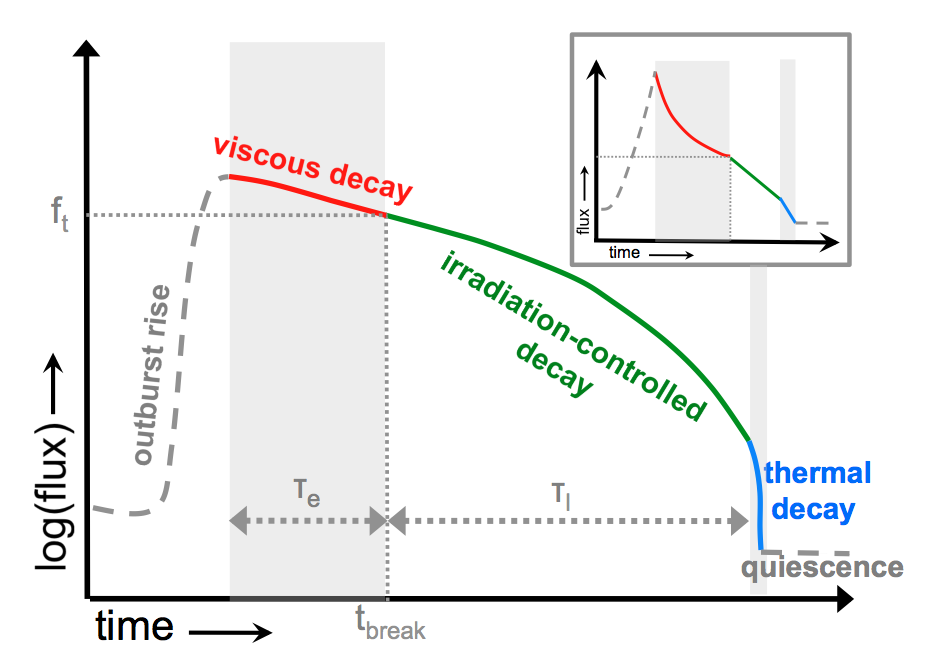}}
\caption{Schematic light-curve for an outburst of a low-mass X-ray binary system. The profile shown corresponds to the light-curve predicted by the (irradiated) disc-instability picture for an outbursting irradiated disc. $\tau_e$ and $\tau_l$ represent the timescales of the exponential (viscous) and linear (irradiation-controlled) decay stages in the light-curve, respectively. The time (and flux) at which the transition between the viscous and exponential stages of the decay occurs (marking the point at which the temperature in the outer disc radii drops below the ionization temperature of hydrogen), are represented by $t_{\rm break}$ and $f_t$, respectively.}
\label{fig:decay_fig_schem}
\end{figure}

\clearpage

\setcounter{table}{0}
\renewcommand{\tablename}{Extended Data Table}
\begin{table}
{
\centering
\caption{ The binary orbital parameters used for our Galactic black-hole low-mass X-ray binary source sample. When no acceptable estimates of distance, black-hole mass $M_1$, or binary mass ratio $q$ are available, known Galactic distributions\cite{tetarenkob2015,oz10} are used.}
\medskip
\label{tab:binaryinfo}
\begin{tabular}{lcccl} 
\hline
		Source Name & distance & $M_1$ & $q$& $P_{\rm orb}$   \\
		 & (kpc) & $(M_{\odot})$ & $(M_2/M_1$) & (hrs)  \\
\hline
XTE J1118+480 & $1.72\pm0.1$ & $7.2\pm0.72$ & $0.024\pm0.009$ & $4.1$\\
	MAXI J1305$-$704&-&-&-&9.74\\
	Swift J1357.2$-$0933&1.5--6.3&$12.4\pm3.6$&-&2.8\\
	GS 1354$-$64&-&-&0.12$\pm$0.04&61.1\\

	4U 1543$-$475&7.5$\pm$0.5&9.4$\pm$2.0&0.25--0.31&26.8\\
	XTE J1550$-$564&4.4$\pm$0.5&10.39$\pm$2.3&0.031--0.037&37.0\\

	XTE J1650$-$500&2.6$\pm$0.7&4.7$\pm$2.2&-&7.7\\

	GRO J1655$-$40&3.2$\pm$0.5&5.4$\pm$0.3&0.38$\pm$0.05&62.9\\
	MAXI J1659$-$152&1.6--8.0&-&-&2.414\\
	GX 339$-$4&8.0$\pm$2.0&-&-&42.1\\

	Swift J1745$-$26&-&-&-&$\leq$21\\

	MAXI J1836$-$194&-&-&-&$<$4.9\\

	XTE J1859+226&8$\pm$3&10.83$\pm$4.67&-&6.6\\
\hline
\end{tabular}\\
}

\end{table}

\clearpage

\begin{table}
{
\centering
\caption{ 
Derived quantities describing the mass transport process in outbursting low-mass X-ray binary accretion discs.
The efficiency of angular-momentum (mass) transport ($\alpha$-viscosity parameter), assuming no mass loss in the hot disc, and related quantities, sampled using our Bayesian Hierarchical Methodology, are presented. The accretion state(s) reached in each outburst are indicated by:  hard (H), intermediate (I), soft (S). For the maximum fraction of the Eddington luminosity, we assume $M_1=10M_{\odot}$$^a$ or $d=8$ kpc$^b$ when one or both of these are unconstrained. }
\medskip
\label{tab:alphainfo}
\footnotesize
\begin{tabular}{lcccccc} 
\hline
		Source Name & Outburst ID&$\tau_e$ &$R_{\rm disc}$& $\alpha_{h}$& Accretion State(s)& Max Eddington\\[-2.5ex]
		&&(days)&($\times 10^{10}$ cm)&&Reached&Fraction\\[-0.5ex]
		
\hline
4U1543$-$475 & 2002& $58.94 \pm 0.42$&${27.91}^{+12.2}_{-10.1}$  & ${0.66}^{+0.16}_{-0.14}$&H,I,S&0.16\\[-0.95ex]
GS1354$-$64 & 2015&${139.69}^{+0.63}_{-0.65}$& ${52.10}^{+18.1}_{-15.3}$ & ${0.362}^{+0.070}_{-0.066}$&H&0.085$^{a,b}$\\[-0.95ex]
GX339$-$4 & 1996-1999& ${167.17}^{+2.12}_{-2.31}$&${37.30}^{+15.6}_{-14.2}$ & ${0.250}^{+0.059}_{-0.056}$&H,I,S&0.011$^a$\\[-0.95ex]
& 2008& ${168.24}^{+5.89}_{-5.81}$&${37.24}^{+15.5}_{-14.1}$ & ${0.247}^{+0.061}_{-0.056}$&H&0.0059$^a$\\[-0.95ex]
& 2009& ${166.88}^{+4.96}_{-4.48}$&${37.33}^{+15.4}_{-14.1}$ & ${0.249}^{+0.060}_{-0.057}$&H&0.0088$^a$\\[-0.95ex]
 & 2013& ${172.37}^{+3.14}_{-3.51}$&${37.29}^{+15.2}_{-14.0}$ & ${0.242}^{+0.058}_{-0.054}$&H&0.014$^a$\\[-0.95ex]
& 2014/2015& ${188.90}^{+0.25}_{-0.23}$&${37.20}^{+15.4}_{-14.0}$ & ${0.222}^{+0.049}_{-0.052}$&H,I,S&0.15$^a$\\[-0.95ex]
MAXIJ1305$-$704 & 2012&${52.90}^{+0.11}_{-0.12}$&${12.72}^{+5.70}_{-4.65}$ & ${0.49}^{+0.11}_{-0.11}$&H,I,S&0.051$^{a,b}$\\[-0.95ex]
MAXIJ1659$-$152 & 2010/2011&${60.69}^{+1.19}_{-1.23}$&${5.494}^{+2.31}_{-2.07}$ & ${0.265}^{+0.059}_{-0.064}$&H,I,S&0.15$^a$\\[-0.95ex]
MAXIJ1836$-$194 & 2011/2012& ${93.09}^{+1.81}_{-2.00}$&${8.838}^{+3.61}_{-3.34}$ & ${0.220}^{+0.049}_{-0.053}$&H&0.11$^{a,b}$\\[-0.95ex]
SwiftJ1357.2$-$0933 & 2011& ${68.31}^{+2.16}_{-2.05}$& ${6.850}^{+3.00}_{-2.46}$ &${0.346}^{+0.067}_{-0.065}$&H&0.0019\\[-0.85ex]
& 2017& ${64.89}^{+3.47}_{-3.68}$&${6.730}^{+2.95}_{-2.42}$&${0.366}_{-0.070}^{+0.066}$&H&0.00038\\[-0.95ex]
SwiftJ174510.8$-$262411 & 2012/2013& ${81.49}^{+1.92}_{-1.86}$ & ${23.49}^{+9.71}_{-8.92}$& ${0.410}^{+0.097}_{-0.091}$&H&1.2$^{a,b}$\\[-0.95ex]
XTEJ1118+480 & 1999/2000& ${85.96}^{+0.55}_{-0.56}$& ${12.94}^{+1.80}_{-2.10}$ & ${0.279}^{+0.017}_{-0.018}$&H&0.0017\\[-0.85ex]
 & 2005& ${79.01}^{+1.29}_{-1.04}$&${12.95}^{+1.78}_{-2.10}$ & ${0.303}^{+0.019}_{-0.021}$&H&0.00047\\[-0.95ex]
XTEJ1550$-$564 & 2000& ${61.78}^{+0.38}_{-0.37}$&${55.97}^{+10.7}_{-9.53}$ & ${0.96}^{+0.15}_{-0.16}$&H,I,S&0.043\\[-0.85ex]
 & 2001& ${61.92}^{+5.04}_{-5.79}$&${56.14}^{+10.7}_{-9.68}$ & ${0.962}^{+0.101}_{-0.089}$&H,I&0.0068\\[-0.95ex]
 & 2001/2002& ${60.38}^{+0.64}_{-0.63}$&${56.06}^{+10.8}_{-9.60}$ & ${0.99}^{+0.15}_{-0.15}$&H&0.013\\[-0.95ex]
& 2003& ${61.89}^{+0.55}_{-0.52}$&${55.99}^{+10.7}_{-9.52}$ & ${0.96}^{+0.15}_{-0.14}$&H&0.015\\[-0.95ex]
XTEJ1650$-$500& 2001/2002& $93.12 \pm 1.26$& ${9.804}^{+4.36}_{-3.56}$ & ${0.185}^{+0.034}_{-0.052}$&H,I,S&0.016\\[-0.95ex]
XTEJ1859+226 & 1999/2000& ${56.61}^{+0.066}_{-0.084}$& ${11.62}^{+5.23}_{-4.27}$ & ${0.505}^{+0.142}_{-0.093}$&H,I,S&0.18\\[-0.5ex]

\hline
\end{tabular}\\
}
\renewcommand*{\arraystretch}{1.0}

\end{table}

\clearpage

\begin{methods}

\subsection{Archival X-ray Data Collection and Reduction}

We have collected all outburst data available since 1996, for each of the 12 systems in our source sample, from the (i) Proportional Counter Array (PCA) aboard the Rossi X-ray Timing Explorer (RXTE), (ii) X-ray Telescope (XRT) aboard the Swift Observatory, (iii) Gas-Slit Camera (GSC) aboard the Monitor of All-sky Image (MAXI) Telescope, (iv) Advanced CCD Imaging Spectrometer (ACIS-S) and High Resolution Camera (HRC-S) aboard the Chandra X-ray Observatory, and (v) European Photon Imaging Camera (EPIC) aboard XMM-Newton.

All X-ray light-curve data from RXTE/PCA was collected from the WATCHDOG project\cite{tetarenkob2015}. These authors compiled all available good pointed PCA observations (i.e., no scans or slews) from the HEASARC archive, for 77 black-hole low-mass X-ray binary sources in the Galaxy, over the entire 16-year RXTE mission.
For each individual source in our sample, we use scripts from the WATCHDOG project, involving the \textit{rex} script within the Heasoft Software Package (http://heasarc.nasa.gov/lheasoft/), to reduce and extract (mission-long) daily time-binned, background-subtracted light curves in the 2--10 keV band, from the PCA Std2 data available on that source in the WATCHDOG database. We have also compiled all available MAXI/GSC data using the WATCHDOG project's online light curve tool\linebreak (http://astro.physics.ualberta.ca/WATCHDOG/download\_data). This tool compiles all the publicly available data from the MAXI archive (http://maxi.riken.jp/top/) in three standard bands ($2-4$, $4-10$, $10-20$ keV), and runs it through the WATCHDOG processing pipeline\cite{tetarenkob2015}. Using this tool, we have extracted (mission-long) daily-time binned, background-subtracted light-curves in the 2--10 keV band, for each individual source (where available).

In addition, we use the Swift/XRT online product builder\cite{evans2009}\linebreak (http://www.swift.ac.uk/user\_objects/index.php) to compile (mission-long) daily time-binned, background-subtracted light-curves in the 2--10 keV band, using all available windowed timing and photon counting mode XRT pointed observations. Lastly, we have collected all available Chandra/ACIS-S, Chandra/HRC-S, and XMM-Newton/EPIC pointed observations from the literature for individual outbursts, where available. We then convert individual count-rates to fluxes in the 2--10 keV band using PIMMS v4.8c \linebreak(http://cxc.harvard.edu/toolkit/pimms.jsp) and the spectral information available in the literature.

\subsection{Conversion from Count-rate to Bolometric Flux}

We use crabs as a baseline unit of flux to calculate approximate count rate equivalences in the 2--10 keV band data from RXTE/PCA, Swift/XRT, and MAXI/GSC. Integration of the now accepted ``canonical'' simple power-law spectrum of the Crab Nebula\cite{tods74}, over the 2--10 keV band, gives us a straightforward method for converting between count rate and flux in this band. Assuming that a source spectrum is Crab-like in nature will cause uncertainty in the computed source flux. However, as it has been found that assuming a Crab-like spectral shape in narrow X-ray energy bands (such as the the 2--10 keV band we make use of here), will produce no more than a 20\% (and typically $<10$\%) error in the source flux for a flat power-law versus a blackbody\cite{tetarenkob2015}, this approach is justified. 

To convert flux in the 2--10 keV band to bolometric flux, we make use of the following bolometric corrections (BCs), estimated for each individual accretion state\cite{migliari2006} occurring during outbursts of black-hole low-mass X-ray binaries; $BC = 5$ (hard state) and a $BC = 1.25$ (soft \& intermediate states). By combining the above discussed bolometric corrections with the
daily accretion state information, obtained from the WATCHDOG project\cite{tetarenkob2015} online Accretion-State-By-Day tool (http://astro.physics.ualberta.ca/WATCHDOG), for each outburst, we are able to compute daily time-binned bolometric light curves.

\subsection{Markov-Chain Monte Carlo (MCMC) Fitting Algorithm}

We make use of a Bayesian approach to estimate the five parameters that describe the shape of an observed light-curve decay profile: the (i) exponential (viscous) decay timescale ($\tau_e$), (ii) linear (irradiation-controlled) decay timescale ($\tau_l$), (iii) X-ray flux of the system at the transition between exponential and linear decay stages ($f_t$), (iv) time after the outburst peak when the transition between exponential and linear decay stages occurs ($t_{\rm break}$), and (v) X-ray flux limit of the exponential decay stage ($f_2$). See Extended Data Figure 1.
Using the emcee python package\cite{for2013}, an implementation of Goodman \& Weare's Affine Invariant MCMC Ensemble Sampler\cite{goodw10}, we apply a MCMC algorithm to simultaneously fit the exponential (viscous) and linear (irradiation-controlled) stages of each decay (as described in the main text and Extended Data Figure 1) where applicable.

Before fitting occurs, secondary maxima and other rebrightening events\cite{kingrit8,menou2000,dubus2001} contaminating the decays are removed by hand. These data are not included in the fits; analysis of these rebrightening events will be presented in a later paper. The removal of these rebrightening events has no effect on the determination of $\alpha$ viscosity from the X-ray light-curves. The remaining data is then fit in logarithmic (bolometric) flux space with our five-parameter analytical model (for details see below). 

The emcee python package runs a modified version of the Metropolis-Hastings Algorithm, in which an ensemble of ``walkers'' simultaneously move through and explore the parameter space. To fit each light-curve, we make use of 50 ``walkers'', 10 times our model dimensions. For the emcee to run optimally, we first appropriately set the initial positions of our ensemble of ``walkers'' in the parameter space. To do so, we make use of pyHarmonySearch\cite{geem2001}, an implementation of the harmony search global optimization algorithm, to perform an initial survey of our parameter space. pyHarmonySearch essentially acts as a less time-consuming version of a brute force grid search method, allowing us to place our ensemble of ``walkers'' in a tight ball around the best guess it finds. This ``best guess'' provides a starting point for the MCMC algorithm.

Prior distributions for each of the five parameters are also set from the results of the pyHarmonySearch of the parameter space. In the case of a well-sampled light-curve (i.e., near-continuous daily data throughout the outburst), a Gaussian prior for each parameter with a mean set by the results of the pyHarmonySearch is used. In the case where only scattered data is available on only a portion of the full decay, wide flat priors (based on expectations from other outbursts of the same source, or outbursts from sources with similar orbital periods) are used for each parameter. 

After initialization, we begin running the MCMC on each light-curve with a 500 step ``burn-in'' phase. Here the ensemble of ``walkers'' are evolved over a series of steps, with the sole purpose of making sure that the initial configuration we have set allows the ``walkers'' to sufficiently explore the parameter space. At the end of the ``burn-in'' phase, if the initial configuration is appropriate for the problem, the ``walkers'' will have ended up in a high probability region, a place in the parameter space where the states of the Markov-chain are more representative of the distribution being sampled. 
After this phase, the MCMC is restarted, with the ``walkers'' starting at the final position they acquired during the ``burn-in'' phase, 
and run until convergence. The number of steps required for convergence is dependent upon the amount of data available and the complexity of the outburst decay profile.

After likelihood maximization is performed, the MCMC algorithm will output the converged solution in the form of posterior distributions of each parameter. We take the best fit result (i.e., the best-fit value along with the upper and lower limits on this value) as the median and 1$\sigma$ (68\%) confidence interval of each posterior distribution, respectively.

\subsection{The Analytical Outburst Decay Model}

 Extended Data Figure 1 shows the predicted characteristic three-stage decay profile shape present in a black-hole low-mass X-ray binary light-curve\cite{kingrit8,dubus1999,dubus2001}.
 
 In the first stage (viscous decay), X-ray irradiation keeps the whole disc in a hot (ionized) state, preventing the formation of a cooling front. As more mass is accreted onto the black-hole than is transferred from the companion at this time, the disc is drained by viscous accretion of matter only, resulting in an exponential-shaped decay profile on the viscous timescale. Eventually, as the mass in the disc, and mass transfer rate, decreases, the dimming X-ray irradiation can no longer keep the outer regions of the disc in the hot (ionized) state and a cooling front forms, behind which the cold matter drastically slows its inward flow. At this point, the system enters the second stage (irradiation controlled decay), during which the propagation of the cooling front is controlled by the decay of the irradiating X-ray flux. The hot (ionized) portion of the disc continues to flow and accrete but gradually shrinks in size, causing a linear-shaped decay profile. Eventually, the mass accretion rate onto the black-hole becomes small enough that X-ray irradiation no longer plays a role. In this third and final stage (thermal decay), the cooling front propagates inward freely on a thermal-viscous timescale, resulting in a steeper linear decay in the light-curve down to the quiescent accretion level.

 The analytical model we use to describe the outburst decay profiles, predicted by the (irradiated) disc-instability picture, in black-hole low-mass X-ray binary light-curves is rooted in the ``classic'' King \& Ritter formalism\cite{kingrit8}. This formalism combines knowledge of the peak X-ray flux and outer radius of the irradiated disc to predict the shape that the decay of an X-ray light curve of a transient low-mass X-ray binary system would follow. 

The temperature of most of the accretion disc in transient low-mass X-ray binaries during outburst is dominated by X-ray heating from the inner accretion region. The X-ray light curve will show an exponential decline if irradiation by the central X-ray source is able to ionize the entire disc, keeping it in the hot (ionized) state and preventing the formation of the cooling front\cite{lasota1}. The X-ray light curve will show a linear decline if irradiation by the central X-ray source is only able to keep a portion of the entire disc in the hot (ionized) state. Then, the central X-ray flux can no longer keep the outer regions of the disc above the hydrogen ionization temperature ($\sim 10^4$ K), and a cooling front will appear and propagate down the disc. As the cooling front cannot move inward on a viscous timescale (i.e., the farthest it can move inward is set by the radius at which $T = 10^4$ K), a linear shaped decline is observed in the light curve. 

By assuming, like many studies of X-ray irradiated discs in close binary systems, an isothermal disc model (i.e., the disc is assumed to be vertically isothermal because it is irradiated, where the central mid-plane temperature is equal to the effective temperature set by the X-ray irradiation flux at the disc surface\cite{dejong1996}), King \& Ritter were able to derive the critical X-ray luminosity for a given disc radius $R_{11}$ (in units of $10^{11}$ cm), 
\begin{equation}
L_{\rm crit} ({\rm BH})=1.7 \times 10^{37} \, R_{11}^2 \, {\rm erg \, s^{-1}},    
\end{equation}
above which the light curve should display an exponential decay shape, and below which the light curve should display a linear decay shape.  

In this formalism, a well sampled light curve (in both time and amplitude) should show a combination of exponential and linear shaped stages in the decay profile. The exponential decay is replaced with a linear decay when the X-ray flux has decreased sufficiently, resulting in a distinct brink (e.g., a break in slope) in the light-curve shape. By deriving analytical expressions for the shape that light curve decays of transient low-mass X-ray binaries systems take, King \& Ritter predicted the timescales of the exponential and linear stages of a decay, the peak mass-accretion rate (and in-turn X-ray luminosity for a given accretion efficiency), and the time at which the exponential decay was replaced by the linear decay.

This approach has since been supported by smooth-particle-hydrodynamics accretion simulations\cite{truss2002} and applied to observations of various classes of X-ray binaries\cite{shab98,powell2007,heinke2015,campana2013,sim06,torres2008} with varied success. However, while the King \& Ritter formalism has, rather coincidentally, been found to agree relatively well with observations, it oversimplifies the physics of the X-ray-irradiated discs to which it is applied\cite{lasota1}. Thus, instead, we make use of a modified version of the King \& Ritter formalism.

In this modified version we (i) include the effects of continuing mass-transfer from the donor star\cite{powell2007,heinke2015}, and (ii) use the disc structure established by Dubus et al.\cite{dubus1999,dubus2001}, where X-ray irradiation affecting black-hole low-mass X-ray binary discs is modelled using a general irradiation law,
\begin{equation}
T_{\rm irr}^4=\frac{C_{\rm irr} L_X}{4 \pi \sigma  R^2}.
\end{equation}
Here, the irradiation parameter $C_{\rm irr}$ is defined as the fraction of the central X-ray luminosity ($L_X=\eta c^2 \dot{M_c}$ for accretion efficiency $\eta$) that heats up the disc. As $C_{\rm irr}$ contains information on the illumination and disc geometry, and the temperature profile of the X-ray irradiation, it effectively parameterizes our ignorance of how these discs are actually irradiated. Physically, $C_{\rm irr}$ controls the timescale of the linear decay stage (and the overall outburst duration), when the transition between decay stages occur, and sets a limit on the amount of mass that can be accreted during the outburst. Stronger irradiation (larger $C_{\rm irr}$) increases the duration of the outburst and thus, the relative amount of matter able to be accreted during an outburst. Consequently, if more matter is accreted during outburst, the following time in quiescence will lengthen, as the disc will require more time to build up again.

Following the procedure outlined in previous work\cite{powell2007,heinke2015}, and instead making use of the general irradiation law defined above, yields the following analytical form for the flux of a black-hole low-mass X-ray binary as a function of time during the exponential (viscous), 
\begin{equation}
f_X=(f_t-f_2)\exp \left( \frac{-(t-t_{\rm break})}{\tau_e} \right)+f_2,
\end{equation}
and linear (irradiation-controlled),
\begin{equation}
f_X=f_t\left(1-\frac{(t-t_{\rm break})}{\tau_l}\right),
\end{equation}
stages of the decay. Here $\tau_e$ and $\tau_l$ are defined as the viscous (exponential) decay timescale in the hot (ionized) zone of the disc and the linear decay timescale, respectively. $f_2=\eta c^2(-\dot{M_2})/4\pi d^2$, is the flux limit of the exponential decay, dependent upon the mass-transfer rate from the companion ($-\dot{M_2}$) and source distance ($d$). $t_{\rm break}$ is defined as the time when the temperature of the outer edge of the disc is just sufficient enough to remain in a hot (ionized) state, and $f_t$ is the corresponding X-ray flux of the system at time $t_{\rm break}$. We perform fits to the flux, as opposed to the luminosity, space to avoid the correlated errors (due to an uncertain distance) that would arise if we were to fit the latter; the uncertain distance (and other parameters) are incorporated below. 

By fitting this model to our sample of observed X-ray light-curves we are able to derive the viscous decay timescales in black-hole low-mass X-ray binaries to range between $\sim50-190$ days, consistent with conclusions of previous works\cite{yan2015}. See Extended Data Table 2 for fit results.

 \subsection{The Bayesian Hierarchical Methodology}
 We quantify angular-momentum (and mass) transport occurring in the irradiated accretion discs present in low-mass X-ray binary systems using the $\alpha$-viscosity parameter. In the current form of the disc instability model, the use of this simple $\alpha$-viscosity parameter results from the inability of current numerical simulations to follow ab initio turbulent transport driven by the magneto-rotational instability on viscous timescales in a global model of the accretion disc.

 This parameter is encoded within the viscous (exponential) stage of the light-curve decay profile. During this first stage of the decay, irradiation of the disc traps it in a hot (ionized) state that only allows a decay of central mass-accretion rate on a viscous timescale,
 \begin{equation}
 \tau_e=\frac{R_{\rm disc}^2}{3 \nu_{\rm KR}}
 \end{equation}
 where $\nu_{\rm KR}$ is the Shakura-Sunyaev viscosity\cite{shak73}, the average value of the kinematic viscosity coefficient near the outer edge of the disc\cite{kingrit8}, and $R_{\rm h,disc}$ is the radius of the hot (ionized) zone of the disc. For Keplerian discs, the Shakura-Sunyaev viscosity is related to the dimensionless viscosity parameter in the hot disc $\alpha_h$ by,
 \begin{equation}
\nu_{\rm KR}=\alpha_h \frac{c_s^2}{\Omega_k},
\end{equation}
where $\Omega_k$ is the Keplarian angular velocity and $c_s$ is the sound speed in a disc (i.e., $c_s\propto T_c^{0.5}$). Thus, using $\Omega_k=(G M_1/R^3)^{1/2}$, the viscous timescale in the disc can be written as a function of the $\alpha$-viscosity parameter in the hot disc $\alpha_h$, compact object mass $M_1$ and accretion disc radius $R_{\rm h,disc}$ such that,
\begin{equation}
\left(\frac{\tau_e}{s}\right)=\left(\frac{G^{0.5} m_H M_{\odot}^{0.5} (10^6)}{3 \gamma k_b T_c}\right) \left( \frac{\alpha_h}{0.1} \right)^{-1}  \left( \frac{M_1}{M_{\odot}} \right)^{0.5}  \left( \frac{R_d}{10^{10} \rm{cm}} \right)^{0.5},
\end{equation}
Because the central midplane temperature of the disc ($T_c$) is only weakly dependent on viscosity and X-ray irradiation in irradiated discs, we can approximate its value as a constant $16300$ K\cite{lasota2015}.

Solving for $\alpha_h$ yields,
\begin{equation}
\left( \frac{\alpha_h}{0.1} \right)=\left(\frac{G^{0.5} m_H M_{\odot}^{0.5} (10^6)}{3 \gamma k_b T_c}\right) \left(\frac{\tau_e}{s}\right)^{-1}  \left( \frac{M_1}{M_{\odot}} \right)^{0.5}  \left( \frac{R_d}{10^{10} \rm{cm}} \right)^{0.5}.
\end{equation}
As the $\alpha$-viscosity parameter in the hot disc ($\alpha_h$) is dependent on parameters characterizing the outburst decay profile of a low-mass X-ray binary (i.e., observed data), as well as the orbital parameters defining the binary system (i.e., parameters that we have prior knowledge of), namely compact object mass and accretion disc radius (which in itself is dependent on the masses of the compact object and companion star in the system and the orbital period), we require a multi-level Bayesian statistical sampling technique to effectively sample $\alpha_h$.

Thus, we have built a Bayesian hierarchical model. A Bayesian hierarchical model is a multi-level statistical model that allows one to estimate a posterior distribution of some quantity by integrating a combination of known prior distributions with observed data. In our case, the (i) established binary orbital parameters (compact object mass, binary mass ratio, orbital period) for a system act as the known priors, and (ii) quantitative outburst decay properties derived from fitting the light curves of a low-mass X-ray binary system with our developed analytical version of the irradiated disc-instability model ($\tau_e$), act as the observed data.

Making use of the emcee python package\cite{for2013} (see above for details), our hierarchical model simultaneously samples $\alpha_h$ for all outbursts of each of the 12 sources in our sample using 240 walkers, 10 times our model dimensions. The model itself has 24 dimensions. 12 of these dimensions correspond to 6 established black-hole mass measurements, 4 known binary mass ratios, and 2 observationally-based Galactic statistical population distributions (the Ozel black-hole mass distribution\cite{oz10} and distribution of binary mass ratios for the dynamically-confirmed stellar-mass black-holes in the Galaxy\cite{tetarenkob2015}). The remaining 12 dimensions correspond to the accretion disc radii for each system. 

Initialization is accomplished by placing our ensemble of ``walkers'' in a tight ball around a ``best guess''. This ``best guess'' corresponds to the best known estimates of the binary parameters (black-hole mass, binary mass ratio, and orbital period) for each system. If a reliable estimate of black-hole mass is not known for a system, the mean of the Ozel mass distribution\cite{oz10} is used. Similarly, if the binary mass ratio is not known for a system, the median of the uniform distribution between the minimum and maximum of the known values of mass ratio for all dynamically confirmed black-holes in the Galaxy\cite{tetarenkob2015} is used.

Our hierarchical model samples accretion disc radii from a uniform distribution between the circularization radius ($R_{\rm circ}$), and the radius of the compact object's Roche lobe ($R_1$) in the system, both of which depend only on the black-hole mass ($M_1$), binary mass ratio ($q$) and orbital period ($P_{\rm orb}$). Initial values of accretion disc radii are set as the median of the uniform distribution between $R_{\rm circ}$ and $R_1$ for each system, calculated using the ``best guess'' for black-hole mass and binary mass ratio (discussed above), and the known orbital period.

The prior distributions for each of the 24 parameters are also set using the ``best guess'' binary orbital parameters for each system. If there exists a constrained measurement of the parameter (i.e., value with uncertainty), a Gaussian prior based on this measurement and its uncertainty is used. If only a range is quoted in the literature for a parameter, a uniform prior is used. The prior distributions for accretion disc radii are taken as the uniform distribution between $R_{\rm circ}$ and $R_1$ for each system.

After initialization, we begin running the emcee sampler on the observed data ($\tau_e$) with a 500 step ``burn in'' phase. After this phase, the emcee sampler is restarted, with the ``walkers'' starting at the final position they ended at in the ``burn-in'' phase, and run until convergence.
Ultimately, the emcee sampler outputs the converged solution in the form of posterior distributions of the $\alpha$-viscosity parameter in the hot disc for each outburst/system. The converged value along with the upper and lower limits on this value is taken as the median and 1$\sigma$ confidence interval of the each posterior distribution, respectively. See Extended Data Table 2.

\subsection{Data Availability}
The data sets generated during and/or analyzed during the current study are available from the corresponding author on reasonable request.

\end{methods}



\nocitemethods{*}
\bibliographystylemethods{naturemag}
\bibliographymethods{methods_refs.bib}

\end{document}